\documentclass[preprintnumbers,prd]{revtex4}
\usepackage{amssymb,amsmath}
\usepackage{graphicx,color}

\newcommand{\non}{\nonumber}
\newcommand{\half}{\tfrac{1}{2}}
\newcommand{\tr}{{\rm tr}}
\newcommand{\UA}{U$_{\rm A}$}
\newcommand{\m}{ \mathcal{M} }

\begin{document}
\title{
\UA(1) breaking and phase transition
in chiral random matrix model} 

\author{T.~Sano$^{a,b}$, H.~Fujii$^a$ and M.~Ohtani$^c$}
\address{%
$^a$Institute of Physics, The University of Tokyo, Tokyo 153-8902,
Japan\\ 
$^b$Department of Physics, The University of Tokyo, Tokyo
113-0033, Japan\\ 
$^c$Physics Department, School of Medicine, Kyorin University, 
Tokyo 181-8611, Japan}

\date{\today}

\begin{abstract}
We propose a chiral random matrix model 
which properly incorporates the flavor-number dependence
of the phase transition owing to the \UA(1) anomaly term.
At finite temperature, the model shows the second-order phase transition
with mean-field critical exponents for two massless flavors, 
while in the case of three massless flavors
the transition turns out to be of the first order. 
The topological susceptibility satisfies the anomalous \UA(1) Ward identity
and decreases gradually with the temperature increased.
\end{abstract}

\maketitle

\section{Introduction}

Chiral symmetry breaking 
manifests itself in the accumulation of the Dirac
modes with zero eigenvalues
through the Banks-Casher relation\cite{BanksC1979}.
In the chiral random matrix (ChRM) theory,
the Dirac operator is restricted to the space of the constant modes
and replaced with a matrix of random entities,
retaining the global symmetries of QCD. 
The ChRM theory has been successful for 
providing a universal framework to
investigate the correlation properties of the low-lying Dirac 
eigenvalues in the so-called epsilon regime in QCD\cite{review,ChRMT}.
The ChRM theory can be also regarded as the simplest schematic model 
for qualitative study of the QCD-like phase diagram from the viewpoint
of chiral symmetry\cite{Halasz:1998qr}.
Effects of the external fields such as the
temperature $T$\cite{Jackson:1995nf, Wettig:1995fg}, 
the quark chemical potential $\mu$
\cite{Stephanov:1996ki,Halasz:1997he,Halasz:1998qr,Stephanov:2006dn,Han:2008xj}
and the Polyakov loop\cite{Stephanov:1996he} have been investigated
within the ChRM models.

The conventional ChRM model predicts a second-order 
phase transition at finite temperature 
$T$\cite{Jackson:1995nf, Wettig:1995fg}, 
irrespective of the number of flavors $N_f$.
This is a shortcoming as a model of QCD. 
Based on the universality argument\cite{Pisarski:1983ms}, 
the chiral transition for $N_f=2$ with the \UA(1) anomaly
is expected to be of the second order
with the O(4) critical exponents,
while the transition becomes of the first order for $N_f \ge 3$.
Even in a mean-field description,
the Ginzburg-Landau effective potential for $N_f =3$
involves the \UA(1) breaking determinant term\cite{KM,tHooft},
which gives rise to a first-order transition.
However, it is unknown so far how to incorporate 
the \UA(1)-breaking determinant term into the ChRM models.

The explicit \UA(1) breaking, or the anomaly, is included
in the ChRM models
by treating $\nu$ exact zero modes with definite chirality,
which should be interpreted as the zero modes associated with
the gauge field configurations of
the topological charge $\nu$. 
The fluctuation of $\nu$ is 
effectively modeled with the instanton ensemble.
The ChRM model supplemented with a Gaussian distribution in $\nu$ 
has been shown to provide a screening 
of the topological charge fluctuation
at zero temperature, 
and thus a resolution of the \UA(1) problem\cite{JanikNPZ97}.   
At finite temperatures, however, this ChRM model
results in an unphysical suppression of 
the topological susceptibility\cite{OhtaniLWH08,LehnerOVW09}.
The resolution of the \UA(1) problem in the ChRM model is then 
limited only at zero temperature.
A modification of the model has been proposed 
in\cite{OhtaniLWH08,LehnerOVW09} to remove the unphysical suppression
of the topological susceptibility at finite temperature.

The ground state is fixed exactly with the saddle point condition
in the thermodynamic limit.
Because the topological charge 
$\nu$ fluctuates around zero with the variance of ${\mathcal O}(V)$,
as dictated by the central limit theorem,  
the nonzero $\nu$ configurations are irrelevant to 
the ground state condition in the ChRM model,
and therefore the \UA(1) breaking seemingly
cannot alter the order of the phase transition at finite temperature.

In this paper we propose a ChRM model involving the \UA(1) breaking
effect, which describes a first-order phase 
transition for $N_f=3$ and at the same time reproduces 
the physical behavior for the topological
susceptibility.
The proposed model consists of the ``near-zero'' modes and
the ``topological zero'' modes 
following Refs.~\cite{JanikNZ97,Jan2}. 
The latters are interpreted as the zero modes
with the right and left chiralities, respectively, 
associated with instantons and anti-instantons. 
We introduce number distributions of the topological zero modes
based on the instanton gas picture. 
Summing over the number of the topological zero modes in the partition function,
we can derive the \UA(1) breaking interaction.
Furthermore, this model satisfies the anomalous \UA(1)
Ward identity\cite{Crewther77}\cite{JanikNPZ97}
and predicts physical temperature dependence
for the topological susceptibility\cite{OhtaniLWH08,LehnerOVW09}.

Interplay between the topological zero modes and the near-zero
modes is studied in Ref.~\cite{Nowak:1989gw}
to derive the instanton-induced interactions
in the coarse-grained instanton-liquid model.
The 'topological' zero modes are also introduced
in Ref.~\cite{JanikNZ97} 
from the determinant interaction 
in the Nambu--Jona-Lasinio (NJL) model in 0+1 dimensions,
but the potential there is unbound in the $N_f=3$ case.

This paper is organized as follows. After reviewing the conventional
ChRM model in the next section, we introduce an extended ChRM model
which involves two kinds of zero modes, the ``near-zero'' modes and
the ``topological zero'' modes, motivated by the instanton gas picture
in subsec.~\ref{sec:IIa}.
Unlike in the conventional ChRM model, the total number of the
topological zero modes is assumed to vary according to the instanton
distribution. 
In subsec.~\ref{sec:IIb} we propose the binomial distribution for the instanton
numbers in a finite space-time volume, and show that
after summing up the topological zero modes the resulting effective
potential has the \UA(1) breaking term and bounded from the below.
General properties of the ground state and fluctuations of the model are
presented for a certain set of parameters and equal quark masses 
in sec.~\ref{sec:III}. 
Section \ref{sec:IV} is devoted to discussions and summary.

\section{Chiral Random Matrix Model}
\label{sec:II}

The QCD partition function with $N_f$ quark flavors of mass $m_f$ 
is written as
\begin{align}
Z_\theta^{\rm QCD}= 
\sum_{\nu =-\infty}^{\infty} e^{i\nu \theta} Z_\nu^{\rm QCD}
=
\sum_{\nu=-\infty}^{\infty}
e^{i\nu \theta} \left\langle \prod_{f=1}^{N_f}\det(D+m_f) \right\rangle_\nu
\; ,
\end{align}
where $\langle \cdots \rangle_\nu$ denotes the average over the gauge field
configurations of fixed topological charge $\nu$, and the $\theta$
parameter has been introduced.

The ChRM model is defined for fixed $\nu$
conventionally as\cite{Jackson:1995nf}
\begin{align}
Z_\nu = \int dW e^{-N\Sigma^2 \tr W^\dagger W} \prod_{f=1}^{N_f} \det(D+m_f) 
\; ,
\end{align}
where the Dirac operator has been replaced with an anti-hermite 
matrix of constant modes
\begin{align}
D=\left ( \begin{array}{cc} 0 & iW+it{\mathbf{1}_{N-|\nu|/2}}
\\ iW^\dagger+it{\mathbf{1}_{N-|\nu|/2}} & 0 \end{array}
\right )
\end{align}
with an 
$(N+\nu/2) \times (N-\nu/2)$ random complex matrix $W$ 
in the chiral representation 
$\gamma_5 ={\rm diag}(1_{N+\nu/2},-1_{N-\nu/2})$. 
Here we have introduced the effective temperature $t$
as a deterministic part in the Dirac operator.
It may be interpreted as the lowest Matsubara frequency
$t = \pi T $.
It is readily shown that the matrix $D$ has $|\nu|$ exact zero
eigenvalues with definite chirality, 
which are interpreted as the exact zero modes associated with
the topological number $\nu$.
The total number of modes $2N$ is finite and proportional
to the space-time volume $V$; $N/V={\cal O}(1)$.

After bosonization of this model
we obtain the effective potential $\Omega_\nu(S)$ on the chiral
manifold;
\begin{align}
Z_\nu
=&\int dS e^{-N\Sigma^2\text{tr}(SS^\dagger)}
\det \left (
(S+m_f)(S^\dagger+m_f )+t^2
\right )^{N-\tfrac{|\nu|}{2}}
\times
\begin{cases}
\det(S+m_f)^{\nu} & (\nu \ge 0)\\
\det(S^\dagger+m_f)^{-\nu} & (\nu<0) \\
\end{cases}
\non \\
\equiv
& \int dS e^{-2N\Omega_\nu(S)}
\; ,
\end{align}
where $S$ is an $N_f \times N_f$ complex matrix of the chiral order
parameter (see next subsection). 
The complete partition function is obtained after summing over $\nu$
\begin{align}
Z_\theta = \sum_{\nu=-2N}^{2N} 
e^{-\frac{\nu^2}{2(2N)\tau}}e^{i\nu \theta}
Z_\nu
.
\label{eq:conv}
\end{align}
Here we need to supplement 
the distribution of $\nu$ characterized by the quenched
topological susceptibility $\tau$, which 
is determined by the pure gluonic dynamics.
Note that the range of the topological charge $\nu$
is limited within $\pm 2N$.

In the thermodynamic limit
$N\to \infty$, the ground state is exactly determined by the saddle
point equation.
Since the fluctuation of $\nu$ scales as
$\nu^2 \sim N$ and nonzero $\nu$ contribution to 
the ground state becomes negligible,
the ground state is given by the saddle point
condition for $\Omega_\nu(S)$ with $\nu=0$,
which is symmetric under U($N_f$)$\times$U($N_f$) for $m_f=0$.
Furthermore, the $N_f$ dependence is factored out in the potential
$\Omega_{0}(S)$ for $S \propto \mathbf{1}_{N_f}$ and 
$m_f=0$,$^{\rm \footnotemark[1]}$%
\footnotetext[1]{
$S \propto \mathbf{1}_{N_f}$ can be derived
automatically from the saddle point condition in the case of equal quark
masses.}
yielding a second-order phase transition irrespective of $N_f$. 
Concerning the fluctuation properties,
the model (\ref{eq:conv}) at zero temperature gives 
a nonzero singlet pseudo-scalar susceptibility $\chi_{\rm ps0}$ 
in the chiral limit, resolving the \UA(1) problem\cite{JanikNPZ97}. 
At finite temperature, however, this leads to an unphysical suppression
of the topological susceptibility as mentioned in Introduction.

\subsection{Model with near- and topological zero modes}
\label{sec:IIa}

Let us consider a variation of
the conventional ChRM model with the instanton gas picture in mind.
An isolated instanton is a localized object accompanying a
right-handed fermion zero mode.
In a dilute system of $N_+$ instantons and $N_-$ anti-instantons,
we expect $N_+$ right-handed and $N_-$ left-handed zero modes 
(even at finite temperature).
In an effective theory at long distances,  effects of the instantons
should be integrated out, which will result in \UA(1)-breaking effective
interactions\cite{tHooft}. 
The fundamental assumption in our modeling is the classification
of the constant modes into 
the ``near-zero'' modes and ``topological zero'' modes\cite{JanikNPZ97}.
We deal with the $2N$ near-zero modes appearing in the conventional
models and include additionally the $N_+ + N_-$ 
topological zero modes which we regard as the modes accompanied by
the instantons.
Distributions of $N_+$ and $N_-$
lead to fluctuations of the topological charge
$\nu=N_+ - N_-$ as well as the total number $N_+ + N_-$.
Eventually we shall sum over $N_+$ and $N_-$ with the mean value of
${\mathcal O}(N)$ and take the thermodynamic
limit $N \to \infty$.

For definite numbers of zero modes,
we write a Gaussian ChRM model as\cite{JanikNZ97} 
\begin{align}
Z^N_{N_+,N_-}=&
\int dA dB dX dY 
e^{-N \Sigma^2 {\rm tr}(AA^\dagger +BB^\dagger +XX^\dagger +YY^\dagger)}
\prod_{f=1}^{N_f} \det(D+m_f) 
\label{eq:fixNpf}
\end{align}
with
\begin{align}
D=
\left(
\begin{array}{cccc}
  0             &iA+it{\mathbf{1}_N} & 0     & iX  \\
  iA^\dagger+it{\mathbf{1}_N} & 0      & iY    & 0   \\
  0             & iY^\dagger & 0 & iB  \\
  iX^\dagger     & 0      & iB^\dagger & 0
\end{array}
\right)
\; .
\end{align}
Here we use the chiral representation in which 
$\gamma_5={\rm diag} (1_N, -1_N, 1_{N_+}, -1_{N_-})$.
Then the matrix $D$, which satisfies
$\{ D,\gamma_5 \}=0$, has a block structure with complex matrices
$A$, $B$, $X$ and $Y$ as shown above.
The $N \times N$ complex matrix $A$ acts on
the near-zero modes, while
the rectangular complex matrix $B$ of size
$N_+ \times N_-$ acts on
the topological zero modes.
The matrices $X \in \mathbb{C}^{N \times N_-}$ and
$Y \in \mathbb{C}^{N \times N_+}$ represent the interactions
between the near-zero modes and the topological zero modes.
In the following we take the quark mass 
$m_f$ of flavor $f$ to be diagonal $\propto
\mathbf{1}_{2N+N_+ + N_-}$
in the space of the zero modes.
Notice that the temperature term $t$
is introduced only for the near-zero modes, while
the topological zero modes are assumed to be
insensitive to the temperature $t$. 
This discrimination is physically legitimate in the instanton gas
picture where each topological zero mode is localized around an
(anti-)instanton and its eigenvalue is not much affected by the
anti-periodic boundary condition in the temporal direction unless
the temporal size $1/T$ becomes the same order of
the typical temporal extent of the instanton or the topological zero
mode.$^{\rm \footnotemark[6]}$
\footnotetext[6]{If $T$ is high enough to modify the nature of the
  topological zero modes, one should deal with the modification of the
  instanton configurations and distributions, too, which is beyond the
  scope of this study.}

The effective potential on the chiral manifold
can be derived from eq.~(\ref{eq:fixNpf}) in a standard manner. First, 
we recast the partition function in the form of the integration over
Grassmann variables of the near-zero modes $\psi$ and
of the topological zero modes $\chi$:
\begin{align}
\det(D+m_f)= 
\int d\psi^\dagger d\psi d\chi^\dagger d\chi
\exp \left[ -
(\psi^\dagger_L, \psi^\dagger_R, \chi^\dagger_L, \chi^\dagger_R)
(D+m_f)
\left (
\begin{array}{c}
\psi_R\\
\psi_L\\
\chi_R\\
\chi_L
\end{array}
\right )  \right] ,
\end{align}
where, for each flavor, $\psi^\dagger_{R,L}$ and $\psi_{R,L}$ have
$N$ components while
$\chi^\dagger_{L{(R)}}$ and $\chi_{R(L)}$ have  $N_{+(-)}$ components.
Then integration over the random matrices $A$, $B$, $X$
and $Y$ with the Gaussian
distributions gives rise to the four-fermion interaction term.
We unfold this term
with the aid of the Hubbard-Stratonovich transformation
as
\begin{align}
&\exp \left[ \frac{1}{N\Sigma^2}
(\psi_L^{\dagger f} \psi_R^g + \chi_L^{\dagger f}\chi_R^g)
(\psi_R^{\dagger g}\psi_L^f + \chi_R^{\dagger g} \chi_L^f)
 \right] 
\non\\
=& \int dS \exp \left[
 -(\psi_L^{\dagger f} \psi_R^g + \chi_L^{\dagger f}\chi_R^g)S_{fg}
 -(\psi_R^{\dagger f}\psi_L^g + \chi_R^{\dagger f} \chi_L^g)S^\dagger_{fg}
 \right] 
\exp \left [ -N\Sigma^2 {\rm tr} S^\dagger S \right ]
\end{align}
with an ${N_f \times N_f}$ complex matrix $S$. 
Here summation over the flavor indices 
$f$ and $g=1,\cdots ,N_f$ should be understood,
and the zero-mode sum is also
implicit in (e.g.) 
$\psi_L^{\dagger f} \psi_R^g \equiv
\sum\limits_{i}\psi^{\dagger f}_{Li} \psi_{Ri}^g$.
This transformation allows us to perform the Grassmann integration.
We thus obtain the desired form for the partition function:
\begin{align}
Z^N_{N_+,N_-}=&\int dS \; e^{-N\Sigma^2 \text{tr} S^\dagger S}
\; \det {}^N \left [ (S+\m)(S^\dagger+\m^\dagger) +t^2 \right ]
\; \det {}^{N_+} (S+\m) \; \det{}^{N_-} (S^\dagger + \m^\dagger) 
 .
\label{eq:fixedia}
\end{align}
The mass matrix $\m=\text{diag}(m_1,\cdots, m_{N_f})$ 
will be taken as
a general $N_f \times N_f$ matrix of the source field 
later in sec.~\ref{sec:III}.

In the chiral limit $\m=0$, the
integrand of (\ref{eq:fixedia}) is invariant under 
$\text{SU}(N_f)\times \text{SU}(N_f)$ transformation,
$S\to USV^{-1}$ with $U, V \in \text{SU}(N_f)$,
in addition to U$_{\rm V}$(1). 
On the other hand, under \UA(1) transformation,
$S\to e^{2i\theta}S$, it acquires a phase factor
$e^{2 i N_f \theta \nu}$ due to the difference $\nu \equiv N_+ - N_-$
in the powers of the determinants. In other words,
the nonzero $\nu$ breaks the \UA(1) symmetry explicitly 
down to $\Bbb Z_{2\! N_f}$.

\subsection{Distributions of topological zero modes} 
\label{sec:IIb}

The complete partition function is obtained after 
summing $Z_{N_+,N_-}^N$ over the `instanton' numbers $N_+$ and $N_-$
with an appropriate weight reflecting the pure gluon 
dynamics.$^{\rm \footnotemark[2]}$%
\footnotetext[2]{We keep here the the number of near-zero modes $2N$ fixed.
In Ref.~\cite{Jan2}, 
the total number of the zero modes, $2N + N_+ + N_-$
was fixed.}
Although the distributions of $N_+$ and $N_-$ will be correlated
in general, here we assume independent distributions  $P(N_\pm)$ for
$N_+$ and $N_-$,   
{\it i.e.},
\begin{align}
Z_\theta =&
 \sum_{N_+,N_-}
e^{i \nu \theta} P(N_+)P(N_-)Z_{N_+,N_-}^N 
=\int dS e^{-2N \Omega(S;t,m,\theta)}
\; .
\label{eq:ia_sum}
\end{align}

First we consider $P(N_\pm)$ in a dilute instanton gas picture.
For one-instanton configuration,
one may assign a weight $\kappa$ compared with a no-instanton
configuration, and multiply a factor $N \propto V$ taking into account 
the integration over the instanton location.
For a configuration with $N_+$ instantons,
we have a weight
\begin{align}
P_{\rm Po}(N_+)=\frac{1}{N_+ !} (\kappa N)^{N_+}
\end{align}
where $N_+ !$ is the symmetry factor~\cite{tHooft}. The same distribution
is assumed for $N_-$ as well. 
This is nothing but the  Poisson distribution up to a normalization.
The summation with $P_{\rm Po}(N_\pm)$ in eq.~(\ref{eq:ia_sum}) results in
the exponentiation of the determinant term\cite{JanikNZ97}:
\begin{align}
\Omega_{\text{Po}}=&
\half \Sigma^2 {\rm tr}SS^\dagger 
-\half\ln \det \left [ (S+\m)(S^\dagger+\m^\dagger) +t^2 \right ]
-\half \kappa 
[e^{i\theta}  \det (S+\m) + e^{-i\theta}\det (S^\dagger +\m^\dagger)]
.
\label{eq:njlpf}
\end{align}
The same determinant term as in
$\Omega_{\text{Po}}$ is commonly 
incorporated in other effective models as the ``anomaly term''
to break the \UA(1) symmetry.$^{\rm \footnotemark[3]}$%
\footnotetext[3]{Although one assumes no mass term in the determinant
interaction conventionally, there is an 
ambiguity\cite{ShifmanVZ}. Our choice of the mass dependence
has been fixed in eq.~(\ref{eq:fixNpf}).}
In the ChRM model, however, this potential is unbound for $N_f=3$.
Indeed, for large $S=\phi {\bf 1}_{N_f}$ 
the term $\det(S+\m) \sim \phi^3$
dominates over the other terms in $\Omega_{\text{Po}}$.

It should be noticed here that the fermion coupling distorts the
$N_\pm$ distribution itself. 
With including the determinant term of the topological zero modes in 
eq.~(\ref{eq:fixedia}), the effective distribution for
$N_+$ reads
\begin{align}
\widetilde P_{\rm Po}(N_+)=\frac{1}{N_+ !} 
\left (\kappa N d\right )^{N_+}
\end{align}
with $d=|\det(S+\m)|$, and similarly for $N_-$. 
We find that the average value of $N_\pm$ increases
indefinitely with increasing $d$ as 
$\left < N_\pm \right > = \kappa N d$. 
However, 
the possibility of infinitely many topological zero modes $N_\pm$
should be avoided in the ChRM model
as a low-energy effective theory within a finite volume $V\propto N$.

Here we regularize the distribution by setting explicitly a maximum
value of ${\cal O}(N)$ for $N_\pm$.
We split the finite space-time volume into $\gamma N$ cells with
$\gamma$ being a constant of ${\cal O}(1)$, and assign a probability
$p$ for a cell to be occupied by a single `(anti-)instanton' 
and $(1-p)$ for a cell unoccupied. 
We exclude the possibility of the double occupation of a cell, which is
justified by a repulsion between the instantons.
Just like in the lattice gas model in statistical mechanics,
this assumption results in the binomial distributions for $N_\pm$:
\begin{align}
P(N_\pm)=
\left (\begin{array}{c}
\gamma N \\ N_{\pm}
\end{array}
\right ) 
\; p^{N_\pm} (1-p)^{\gamma N-N_\pm}
.
\label{eq:bino}
\end{align}
For a small $p$ and a large $\gamma N$, the binomial distribution
$P(N_\pm)$ is accurately approximated with the Poisson 
distribution with the mean $\gamma N p$. 
But it cannot for a large $p$.
The binomial distribution provides a stringent upper bound $\gamma N$ for
the number of topological zero modes $N_\pm$,
in contrast to the Poisson distribution.
The corresponding effective potential for $S$ is found to be
\begin{align}
\Omega(S;t,m,\theta)=&
\half \Sigma^2 {\rm tr}SS^\dagger 
-\half\ln \det \left [ (S+\m)(S^\dagger+\m^\dagger) +t^2 \right ]
\non \\
&-\half \gamma 
\left [\ln \left(e^{i\theta}  \alpha \det (S+\m)+1 \right)
     + \ln \left(e^{-i\theta}  \alpha \det (S^\dagger +\m^\dagger)+1\right)
\right ]
\label{eq:thepf}
\end{align}
with $\alpha=p/(1-p)$.
This is the ChRM model that we propose and analyze in this paper.

The effective potential~(\ref{eq:thepf}) is bounded from the below
by the $\text{tr}SS^\dagger$ term in contrast to $\Omega_{\rm Po}$
in eq.~(\ref{eq:njlpf}).
The anomaly terms are accommodated
under the logarithms in the square bracket.
Moreover, for small value of $|\alpha \det (S+m)|$,
it reduces to the potential (\ref{eq:njlpf}) with $\kappa=\gamma \alpha$. 
On the other hand, 
the Poisson approximation fails 
when $\left < N_\pm \right >$ becomes ${\cal O}(\gamma N)$.
We stress here again the fact that the distributions $P(N_\pm)$ is
deformed in the presence of the coupling with the topological zero
modes as
\begin{align}
\widetilde P(N_+) =
\left (\begin{array}{c}
\gamma N \\ N_{+}
\end{array}
\right ) 
(p d)^{N_+} (1-p)^{\gamma N-N_+}
,
\end{align}
which changes the probability $p$ to an effective one
$\tilde p \equiv p d/(p d+1-p)$.
Similarly for $N_-$.
Accordingly, the mean number of the zero modes
is modified to 
$\langle \widetilde{N}_ \pm \rangle=\gamma N \tilde p$.

Two remarks are here in order.
In the conventional models, the total number of the modes is kept
fixed, while the fluctuation of $\nu$ is allowed for resolving the
\UA(1) problem.
On the other hand, in the instanton gas picture, 
both of $N_+$ and $N_-$ are expected to fluctuate naturally.
As stressed in this subsection,
the mean number of the modes depends on the magnitude of  
$S$, which gives an ${\mathcal O}(N)$ effect.
The anomaly term appearing in the effective potential in turn
affects the saddle point value of $S$.
This point is essential to the first-order transition for
$N_f=3$, which is overlooked in the earlier works.

Secondly, in eq.~(\ref{eq:thepf})
one needs to introduce a dimensionful scale to compensate 
the dimension of determinant if the dimension of mass is assigned to
$\m$ as well as $S$.
Here we leave all the quantities dimensionless for demonstration of general
feature of the model.

\subsection{$\theta$ dependence}

The variance of the topological charge $\nu = N_+ - N_-$
for the binomial distribution is computed
as $2N \tau = 2N \gamma p (1-p)$, where $\tau$ is the quenched
topological susceptibility. In the presence of the fermion coupling, 
this susceptibility will be replaced with
\begin{align}
\tilde \tau=\gamma \tilde p (1-\tilde p)
=
\gamma \frac{\alpha d}
            {(\alpha d+1 )^2}
\; .
\label{eq:tildetau}
\end{align}
We can confirm this fact also by rewriting
the anomaly term in eq.~(\ref{eq:thepf}) as
\begin{equation}
-\frac{1}{2}\gamma\ln\left[1+|\alpha\det(S+{\cal M})|^2
+2|\alpha\det(S+{\cal M})|\cos\left(
\theta-\frac{i}{2} \ln 
\frac{\text{det}(S+\m)}{\text{det}(S^\dagger + \m^\dagger)}
\right)\right]\ ,
\label{eq:largeNcform}
\end{equation}
from which we find again the replaced topological susceptibility $\tilde\tau$
as a coefficient of $\theta^2$.
Such a series in $\theta$ of the anomaly term 
gives a connection to the general arguments on 
the $\eta'$ in the $1/N_{\rm c}$ expansion \cite{Witten80}.

To make a clear connection with the conventional model, 
let us ignore for a moment the fluctuation of the total number
$N_+ + N_-$ and apply the Gaussian approximation for the $\nu$
distribution.
We then obtain the partition function as
\begin{align}
Z_\theta =& \int_{-\infty}^{\infty}d\nu e^{-\tfrac{\nu^2}{2(2N)\tau}}
\;  e^{i\nu\theta} \; Z_{N_+,N_-}^N
\non \\
=&
\text{const.}\int dS \; e^{-N\Sigma^2 \text{tr} S^\dagger S}
\; \det{}^N \left[ (S+\m)(S^\dagger+\m^\dagger)+t^2 \right ] 
\non \\
& \qquad \times
\; \det{}^{(N_+ +N_-)/2}\left [(S+\m)(S^\dagger + \m^\dagger)\right ]
\exp \left [ -N\tau
\left ( 
\theta-\frac{i}{2} \ln 
\frac{\text{det}(S+\m)}{\text{det}(S^\dagger + \m^\dagger)}
\right )^2
\right ] 
\; .
\label{eq:GaussSum}
\end{align}
This is almost identical with the model 
discussed in Ref.~\cite{OhtaniLWH08,LehnerOVW09}, which reproduces the
screening of the (unquenched) topological susceptibility 
(see eq.~(\ref{eq:topsus2}))
as measured on a lattice at finite temperatures.
However, this model~(\ref{eq:GaussSum}) fails to describe a
first-order phase transition at finite temperature for $N_f$=3.
In fact, the anomaly term appears only as a phase 
in (\ref{eq:GaussSum}) in contrast to
(\ref{eq:largeNcform}) and drops out when we determine the magnitude of
$S=S^\dagger$ in the ground state (with $\m = \m^\dagger$). 
The variation of $N_+ + N_-$ is essential for the anomaly term to affect
the saddle point condition, and thus the order of the phase transition.

\section{Ground state and fluctuations}
\label{sec:III}

In this section we shall study the ground state properties
of the system with equal mass, 
$\m=m \mathbf{1}_{N_f}$, for simplicity.
Setting $S = \phi {\bf 1}_{N_f}$
with real $\phi$ and with $\theta=0$,
we obtain a simple form of the grand potential:
\begin{align}\label{eq:gtpot}
\Omega (\phi; t,m)=&
 \half N_f\Sigma^2 \phi^2
- \half N_f \ln \left[ (\phi + m)^2 + t^2 \right]
- \gamma \ln \left | \alpha (\phi + m)^{N_f} +1 \right |
\; .
\end{align}
The factor $N_f$ cannot be factored out in the potential $\Omega$
because of the anomaly term here.

\subsection{Chiral phase transition: ground state}

In the thermodynamic limit, $N\to \infty$,
the ground state can be 
analyzed with the solution of the saddle point equation
\begin{align}
\Sigma^2 \phi -\frac{\phi+m}{(\phi+m)^2+t^2}
-\gamma \frac{\alpha (\phi+m)^{N_f-1}}{\alpha(\phi+m)^{N_f}+1} =0 
\; .
\label{eq:saddleq}
\end{align}
The scalar quark condensate is related to the solution
$\phi$ as
\begin{align}
\left<\bar{\psi}\psi \right> 
= -\frac{1}{2N N_f}\frac{\partial}{\partial m} \ln Z(m)
= - \Sigma^2 \phi
.
\end{align}
Without the anomaly term $\alpha=0$, eq.~(\ref{eq:saddleq})
recovers the flavor-independent gap equation of the conventional model,
which has the solution $\phi_0^2=\Sigma^{-2}-t^2$
in the chiral limit. 
One can estimate the effect of small $m$ and $\alpha$
on $\phi^2$ at the leading order as
\begin{align}
\phi^2 & \sim
 \phi_0^2 -\frac{m}{\phi_0} (\Sigma^{-2}-2 t^2) 
+\frac{\alpha \gamma}{\Sigma^4}\phi_0^{N_f-2}\; . 
\end{align}
At lower temperatures $2t^2< \Sigma^{-2}$ 
the value of the condensate is decreased by the quark mass term,
on the contrary to our intuition.
This is because the leading order term of $m$ appears in a
combination $-2m\phi/(\phi^2 +t^2)$ in the potential, 
which simplifies to $-2m/\phi$ for $t=0$, favoring smaller
$\phi$ for the potential to be more stabilized.

There is no chiral symmetry for $N_f=1$ because of the anomaly term.
Let us discuss phenomenologically  interesting cases, $N_f=$ 2 and 3.

\subsubsection{$N_f=2$}

Expanding the potential $\Omega$ around $\phi=0$, we find
\begin{align}
\Omega=c_0+c_2\phi^2 +c_4 \phi^4 -h \phi + {\cal O}(\phi^6,\phi^3 m,m^2)
\end{align}
with $c_0=-\ln t^2$, $c_2=\Sigma^2-\alpha \gamma - t^{-2}$,
$c_4=(\gamma\alpha^2+t^{-4})/2$ and $h=2(\alpha \gamma +t^{-2})m$.
This is the standard form of the Landau-Ginzburg potential for a
second order phase transition ($m=0$) with the critical temperature $t_c$
\begin{align}
t_c=\frac{1}{\sqrt{\Sigma^2-\alpha \gamma}}
\; .
\end{align}
Inclusion of the anomaly term increases the value of $t_c$. 
The behavior in the vicinity of the critical point
is characterized by the mean-field exponents; 
we find $\beta=1/2$ and $\delta=3$, respectively, 
from the solutions
$\phi^2 = \epsilon
t_c^{-2}/c_4 \propto \epsilon^{2\beta}$
for $\epsilon=(t_c-t)/t_c$
with $0<\epsilon \ll 1$,
and 
$\phi=[m\Sigma^2/(2c_4)]^{1/3}
\propto m^{1/\delta}$
with $m\ne 0$ at $t=t_c$.

\subsubsection{$N_f=3$}

Expanding the potential $\Omega$ around $\phi=0$, we find
\begin{align}
\Omega=c_0+c_2\phi^2 +c_3 \phi^3 +c_4 \phi^4 -h \phi +{\cal O}(\phi^6,\phi^2 m, m^2)
\end{align}
with 
$c_0= - (3/2)\ln t^2$, 
$c_2=(3/2)(\Sigma^2 - t^{-2})$,
$c_3= - \alpha \gamma$,
$c_4=(3/4)t^{-4}$ and $h=3t^{-2} m$. The \UA(1) symmetry is explicitly
broken even for $m=0$ due to the anomalous $\phi^3$ term, which leads
to a first-order phase transition. 

\begin{figure}[tb]
\begin{center}
\includegraphics[height=8cm,angle=270]{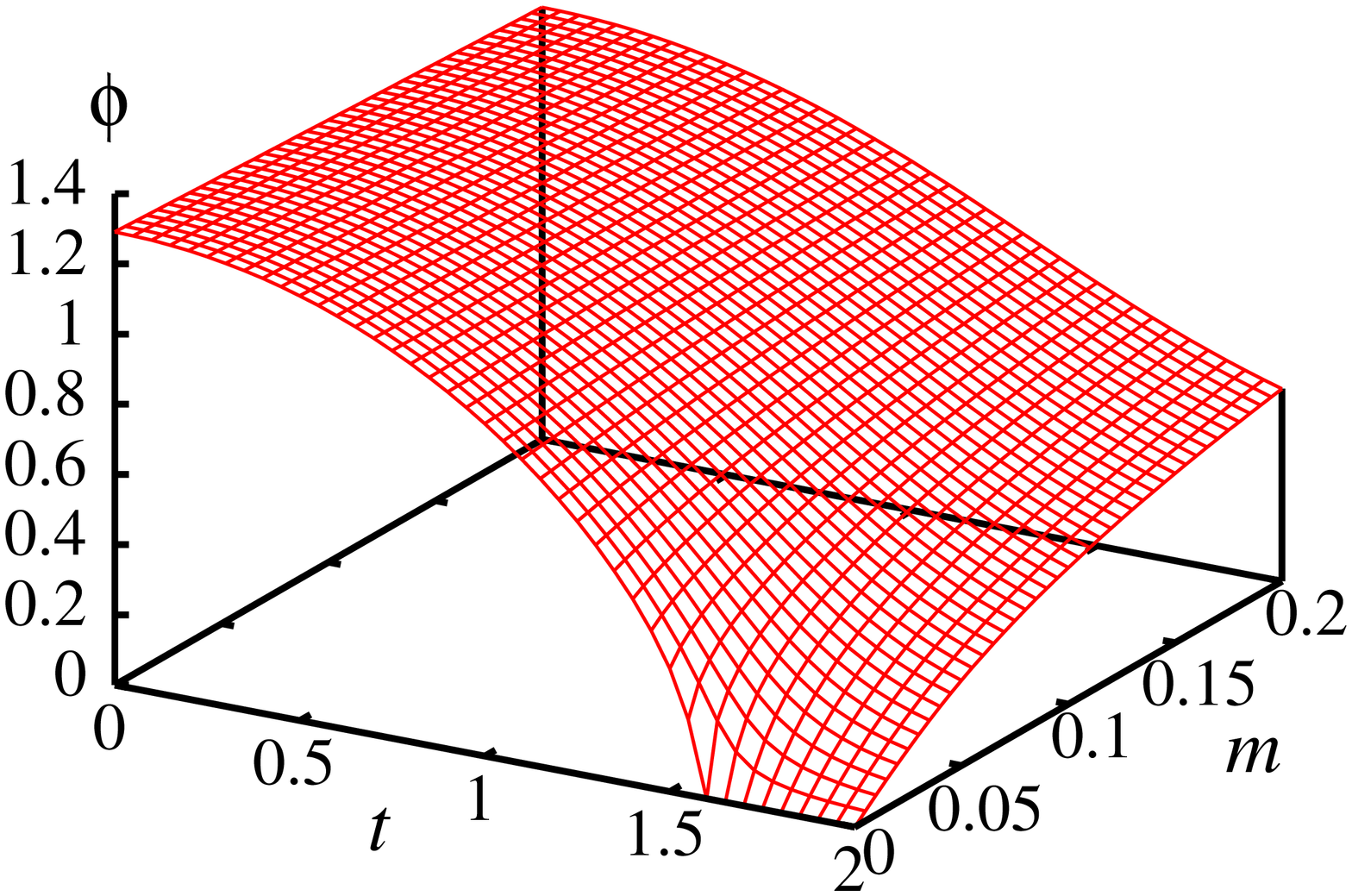}
\hspace{2.5em}
\includegraphics[height=8cm,angle=270]{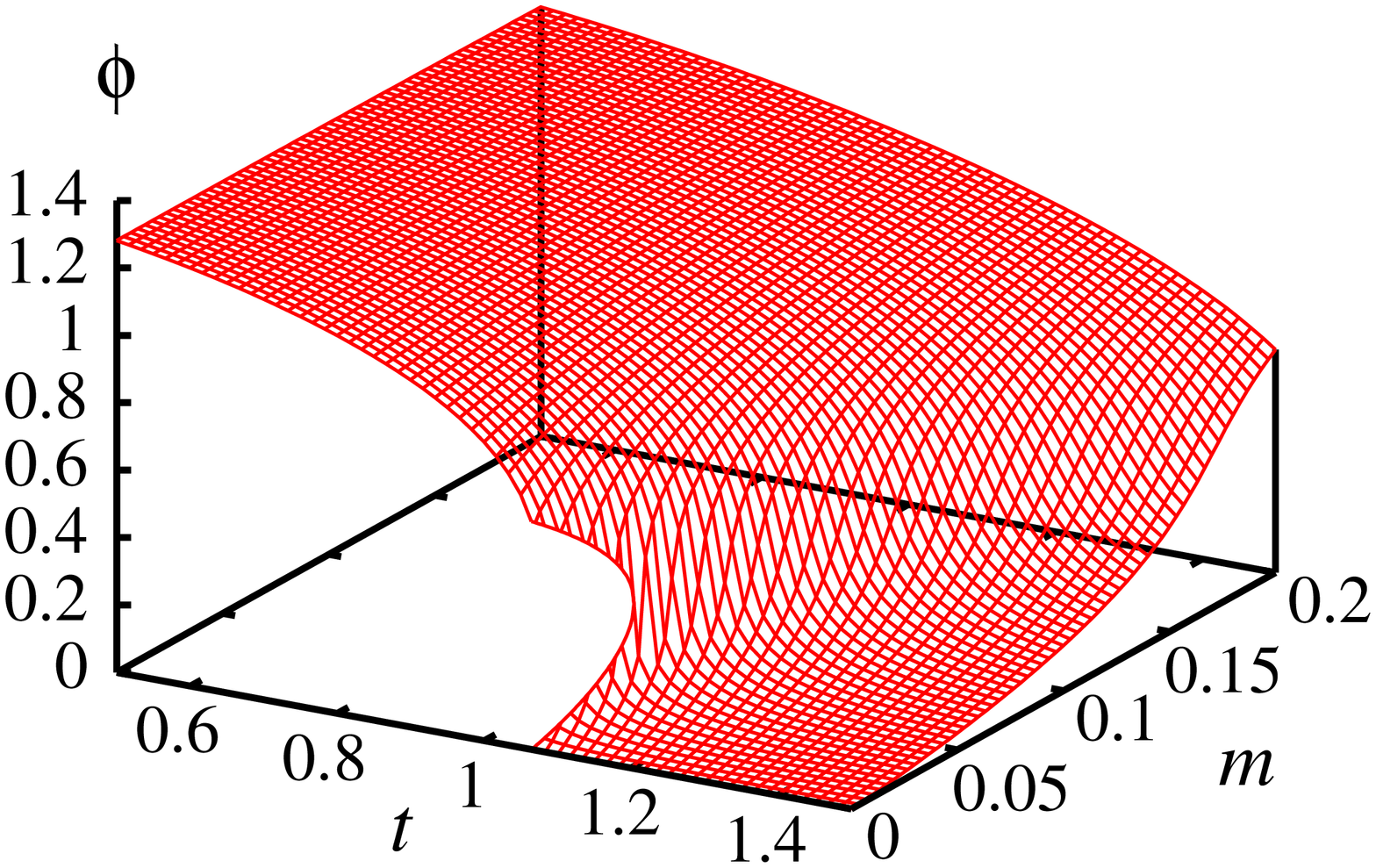}
 \caption{Chiral condensate $\phi$
as a function of $t$ and $m$ for $N_f$=2 (left) and 3 (right).
\label{fig:cond}}
\end{center}
\end{figure}

In Fig.~\ref{fig:cond}, 
we display the chiral condensate $\phi$ as a function
of the temperature $t$ and the quark mass $m$.
For numerical demonstration, we have chosen 
the parameters as $\Sigma=1$, $\gamma=2$ and $\alpha=0.3$ in this paper. 
We clearly see the second order transition for $N_f=2$,
while the first order transition for $N_f=3$ in the chiral limit $m=0$.
In $N_f=3$ case, as we increase the current quark mass
$m$, we find a terminating point of the
first order line at $m_c=0.0265$.

For $N_f \ge 4$, the anomaly term only affects the coefficients of
$\phi^n$ ($n \ge N_f$) in the series expansion of the potential. 

\subsection{Mesonic masses}

Once the ground state acquires the 
nonzero expectation value 
$ S = \phi \mathbf{1}_{N_f} \ne 0$,
the symmetry is spontaneously 
broken down to  U$_{\rm V}(N_f)$; 
$S \to U S U^{-1}$.
Degeneracy of the vacua in the chiral limit
dictates the massless fluctuations
$S=\phi e^{i \pi^a \lambda^a/(\sqrt{2}\phi)} \sim
\phi + i \pi^a \lambda^a/\sqrt{2}$
corresponding to the generators of the broken symmetry.

To find the mesonic masses, 
we use the parametrization
$S=\phi+\lambda^a (\sigma^a+i\pi^a)/\sqrt{2}$
with $\sigma^a, \pi^a \in \mathbb{R}$
and the U($N$) generators $\lambda^a$
normalized as $\text{tr}[\lambda^a \lambda^b]=2\delta^{ab}$
($a=0, \cdots, N_f^2-1$).
Using a formula for a matrix $X$ with a small parameter $\epsilon$, 
$\det (1 + \epsilon X)=
1 + \epsilon\text{tr}X
+\tfrac{1}{2}\epsilon^2 [(\text{tr}X)^2 -\text{tr}X^2]+{\cal O}(\epsilon^3)$,
we can easily expand $\Omega$ (\ref{eq:thepf}) around the 
saddle point solution to define the massses with
$\Omega = \Omega_0  + \tfrac{1}{2}M_{{\rm s}a}^2 \sigma^{a2}
+ \frac{1}{2}M_{{\rm ps}a}^2 \pi^{a2} + \cdots$.

The flavor non-singlet masses in the scalar and
pseudo-scalar channels, respectively, are found to be
\begin{align}
M_{\rm s}^2 =&
 \Sigma^2 - \frac{1}{(\phi+m)^2 +t ^2} 
+ \frac{2(\phi+m)^2}{[(\phi+m)^2 +t ^2]^2} 
+\gamma \frac{\alpha(\phi+m)^{N_f-2}}{\alpha(\phi+m)^{N_f}+1}
\; ,
\\
M_{\rm ps}^2 =&
 \Sigma^2 - \frac{1}{(\phi+m)^2 +t ^2} 
-\gamma \frac{\alpha(\phi+m)^{N_f-2}}{\alpha(\phi+m)^{N_f}+1}
\; .
\end{align}
With the saddle point equation, we can re-express the $\pi$ mass
as
\begin{align}
M_{\rm ps}^2=\frac{\Sigma^2 m}{\phi+m}
.
\end{align}
In the chiral limit, we have the massless $\pi$.
With the explicit breaking of $m$, we find a relation reminiscent
of the Gell-Mann-Oakes-Renner, 
\begin{align}
(\phi+m)^2 M_{\rm ps}^2 =m\Sigma^2 (\phi+m) \sim 
-m \langle \bar \psi \psi \rangle
,
\end{align}
if we identify $\phi+m$ as the pion decay constant $f_\pi$.
Interestingly, there is a mass hierarchy,
$M_{\rm s}^2:\Sigma^2:M_{\rm ps}^2=2:1:0$,
in the chiral limit at zero
temperature $t=0$.

On the other hand, the flavor singlet masses 
for the scalar and pseudo-scalar singlet channels
have an additional
contribution from the anomaly term as
\begin{align}
M_{\rm s0}^2 =&
M_{\rm s}^2  -\Delta M_0^2
\; ,
\label{eq:s0mass}
\\
M_{\rm ps0}^2 =&
M_{\rm ps}^2 +\Delta M_0^2
\label{eq:ps0mass}
\end{align}
with 
\begin{align}
\Delta M_0^2 
&\equiv 
N_f \gamma\frac{\alpha(\phi+m)^{N_f-2}}{ [ \alpha(\phi+m)^{N_f}+1 ]^2} 
\nonumber
\\
&= N_f \frac{\tilde \tau }{(\phi+m)^2}
\; .
\end{align}
The would-be Nambu-Goldstone (NG) mode in the singlet channel becomes massive due to the
coupling to the \UA(1) interaction $\Delta M_0^2$. This same effect appears
as a reduction in the scalar singlet channel.
This mass gap is related to the (replaced) quenched susceptibility
$\tilde \tau$,
similarly to the Witten-Veneziano formula\cite{Witten80}.
It is interesting to note that
for $N_f \ge 3$
the \UA(1) breaking effect 
disappears $\Delta M_0^2=0$ from the meson masses
when the condensate vanishes in the symmetric
phase in the chiral limit\cite{Lee:1996zy}.
In contrast,
$N_f$-point functions are affected by the \UA(1) breaking term.

\begin{figure}[tb]
\begin{center}
\includegraphics[width=8cm]{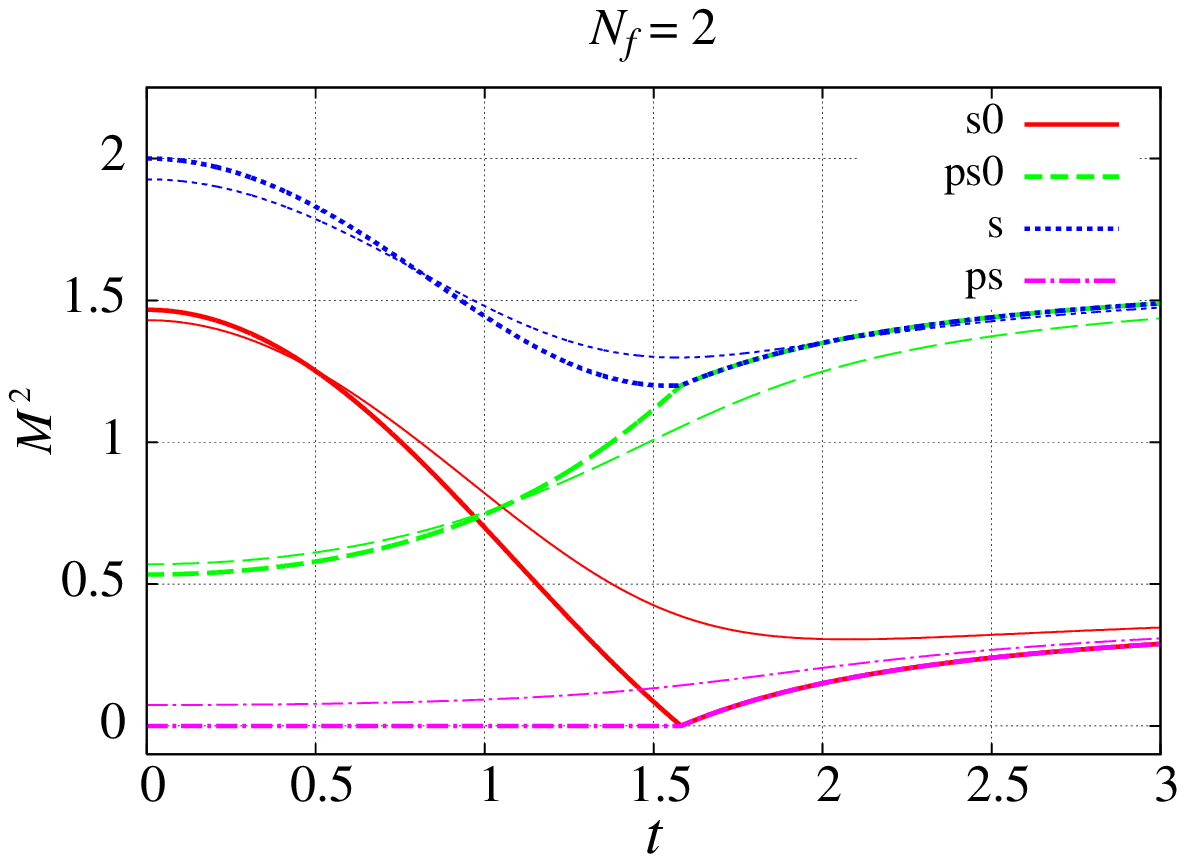}\hspace{2.5em}
\includegraphics[width=8cm]{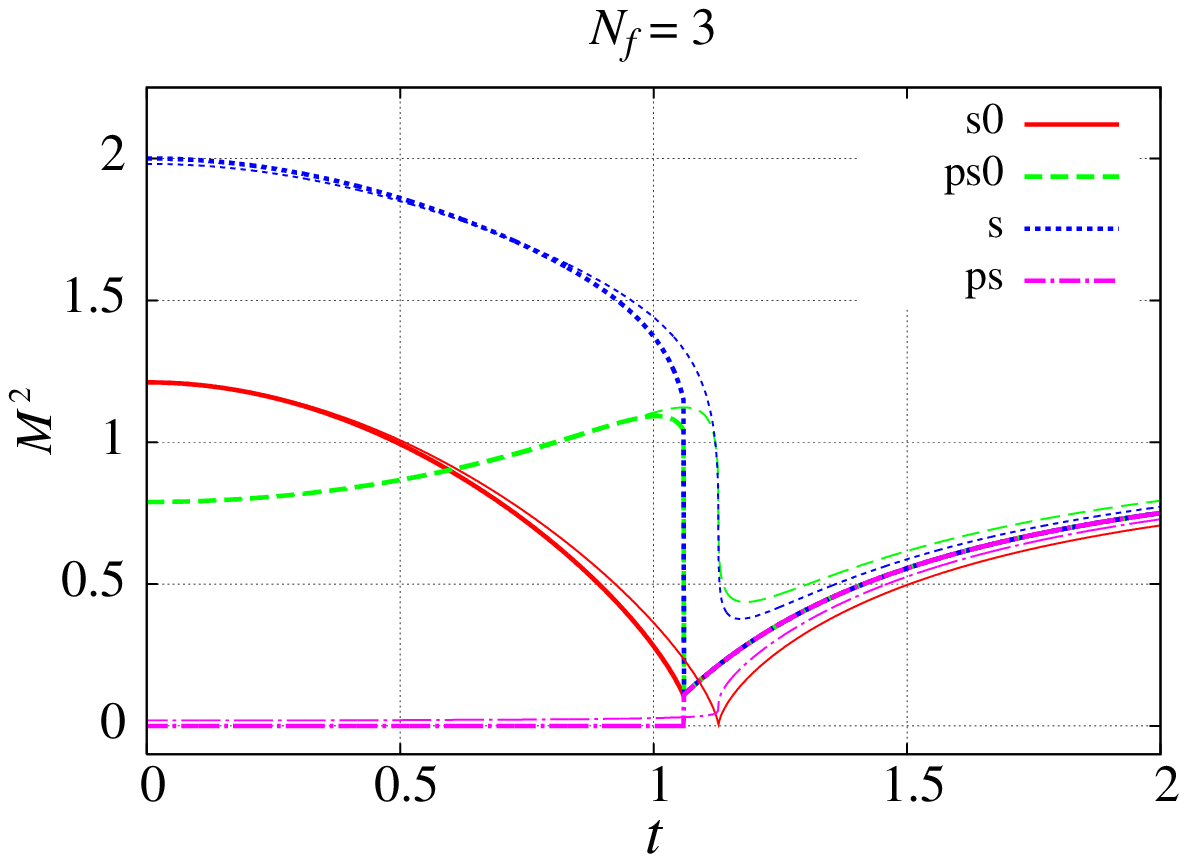}
\caption{Temperature dependence of 
the mesonic masses in the flavor-singlet scalar (s0)
and pseudo-scalar (ps0) channels and in the flavor-nonsinglet
scalar (s) and pseudo-scalar (ps) channels
 in the chiral limit 
($m=0$; thick lines) and with explicit breaking ($m \ne 0$; thin lines)
for $N_f=2$ (left) and $N_f=3$ (right).
As nonzero quark mass, we set $m=0.1$ for $N_f$=2 and
$m=m_c=0.0265$ for $N_f=3$.
}
\label{fig:mass}
\end{center}
\end{figure}

In Fig.~\ref{fig:mass} we show the mesonic masses in the flavor-singlet
scalar and pseudo-scalar channels and in the flavor-nonsinglet scalar
and pseudo-scalar channels. Both for $N_f$=2 and 3, the flavor-singlet
pseudo-scalar meson acquires the nonzero mass via (\ref{eq:ps0mass}).

\subsection{Susceptibilities} 

The scalar and pseudo-scalar
susceptibilities are defined as the responses 
to the external fields $\m= (s^a + ip^a)\lambda^a /\sqrt{2}$ 
in eq.~({\ref{eq:thepf}}):
\begin{align}
\chi_{\rm s}^{ab}=&
 \left. -\frac{\partial^2}{\partial s^a \partial s^b}
\Omega (S(\m);\m)\right |_{\m=m}
\; ,
\nonumber \\
\chi_{\rm ps}^{ab}=&
 \left. -\frac{\partial^2}{\partial p^a \partial p^b}
\Omega(S(\m);\m) \right |_{\m=m}
\; .
\end{align}
In our basis, those susceptibilities are found to be diagonal
$\chi_{\rm s, ps}^{ab} \propto \delta^{ab}$
with a simple form:
\begin{align}
\chi =&
\chi^{(0)} \frac{1}{1-\chi^{(0)}/\Sigma^2}
=
\chi^{(0)} \frac{\Sigma^2}{M^2}
\end{align}
with 
$\chi^{(0)}= \Sigma^2 - M^2$
for each channel.
A susceptibility diverges when the corresponding $M^2$ vanishes
(e.g.) at a critical point where the effective potential 
$\Omega(S)$ has a flat direction.
Especially, in the broken phase, the 
susceptibilities of the non-singlet NG modes are indefinite.

The topological susceptibility, $\chi_{\text{top}}$ is defined 
as the response to the angle $\theta$
\begin{align}
\chi_{\text{top}} \equiv 
 \left .
-\frac{1}{2N}\frac{\partial^2}{\partial \theta^2} \ln Z_\theta
\right |_{\theta=0}\; 
.
\label{eq:topsus}
\end{align}
Using the saddle point solution with small $\theta$,
parametrized as
$S(\theta)=\phi+i\eta_0(\theta)\lambda^0/\sqrt{2}$,
the susceptibility in the mean field approximation
is obtained by differentiating
$\Omega(S(\theta);\theta)$ around $\theta=0$
as 
\begin{align}
\chi_{\rm top} =&
\frac{\partial^2 \Omega}{\partial \theta^2}
-
\left (
\frac{\partial^2 \Omega}{\partial \theta \partial \eta_0}
\right)^2
\left (\frac{\partial^2 \Omega}{\partial \eta_0^2} \right) ^{-1}
\non \\
=&
\tilde \tau
-\frac{N_f\; \tilde \tau^2}{(\phi+m)^2} 
\frac{1}{M_{\rm ps 0}^2}
\non \\
=&
\left [
\frac{1}{ \tilde \tau } +\frac{N_f}{m\Sigma^2 (\phi+m)}
\right]^{-1}
\; ,
\label{eq:topsus2}
\end{align}
or
\begin{align}
\frac{1}{ \chi_{\text{top}} }
= 
\frac{1}{\tilde \tau } + \frac{1}{\tau_m}
\; .
\label{eq:chitop}
\end{align}
Here $\tilde \tau$ is the modified susceptibility defined in
eq.~(\ref{eq:tildetau}) in the previous section. 
As is well known\cite{Leutwyler:1992yt}, 
the most prominent effect of the fermion coupling
is the screening of the topological susceptibility
via the contribution
\begin{align}
\tau_m= \frac{\Sigma^2 m (\phi+m)}{N_f}
=\frac{M_{\rm ps}^2 (\phi+m)^2}{N_f}
\; .
\end{align}
In the quenched limit, $N_f \to 0$, 
the $\chi_{\rm top}$ 
recovers the quenched susceptibility $\tau = \gamma p (1-p)$, 
while in the massless limit
the susceptibility $\chi_{\text{top}}$ is screened completely
to zero.

By changing the variable $S\to \tilde S=Se^{i\theta/N_f}$ under the integral
(\ref{eq:thepf})
and then applying the saddle point approximation,
we find $\Omega(t,m,\theta)=\Omega(t, m e^{i\theta/N_f},0)$.
Here 
the $\theta$  appears only in the combination with the mass $m$.
In this latter form, 
the source term becomes a linear combination
of the scalar and pseudo-scalar sources in the flavor-singlet channel:
$\mathcal{M}=m e^{i\theta/N_f}=(s_0 +i p_0)\lambda^0/\sqrt{2}$,
and the \UA(1) relation for the topological susceptibility\cite{Crewther77} is
immediately derived as 
\begin{align}
\chi_{\text{top}} &
=
- \frac{1}{2N}
\left(
  \frac{\partial^2\ln Z_\theta}{\partial p_0^2}
\left .  
\left  (
\frac{\partial p_0} {\partial \theta}
\right )^2
+
  \frac{\partial \ln Z_\theta}{\partial s_0}
  \frac{\partial^2 s_0}{\partial \theta^2}
\right)\right |_{\theta=0}
\non \\
&=
-\frac{m^2}{N_f}\chi_{\text{ps}0}
-\frac{m}{N_f}\langle \bar \psi \psi \rangle
\; .
\end{align}
We have seen that the pseudo-scalar meson in the flavor singlet channel
has nonzero mass (\ref{eq:ps0mass})
because of the \UA(1) breaking term, 
and accordingly the pseudo-scalar singlet susceptibility remains finite
in the broken phase in the chiral limit. Thus for the small but nonzero
quark mass $m$, the topological susceptibility $\chi_{\rm top}$
decreases following the chiral condensate 
$\langle \bar \psi \psi \rangle \sim \phi$ with increasing temperature $t$.

\begin{figure}[tb]
\begin{center}
\includegraphics[width=8cm]{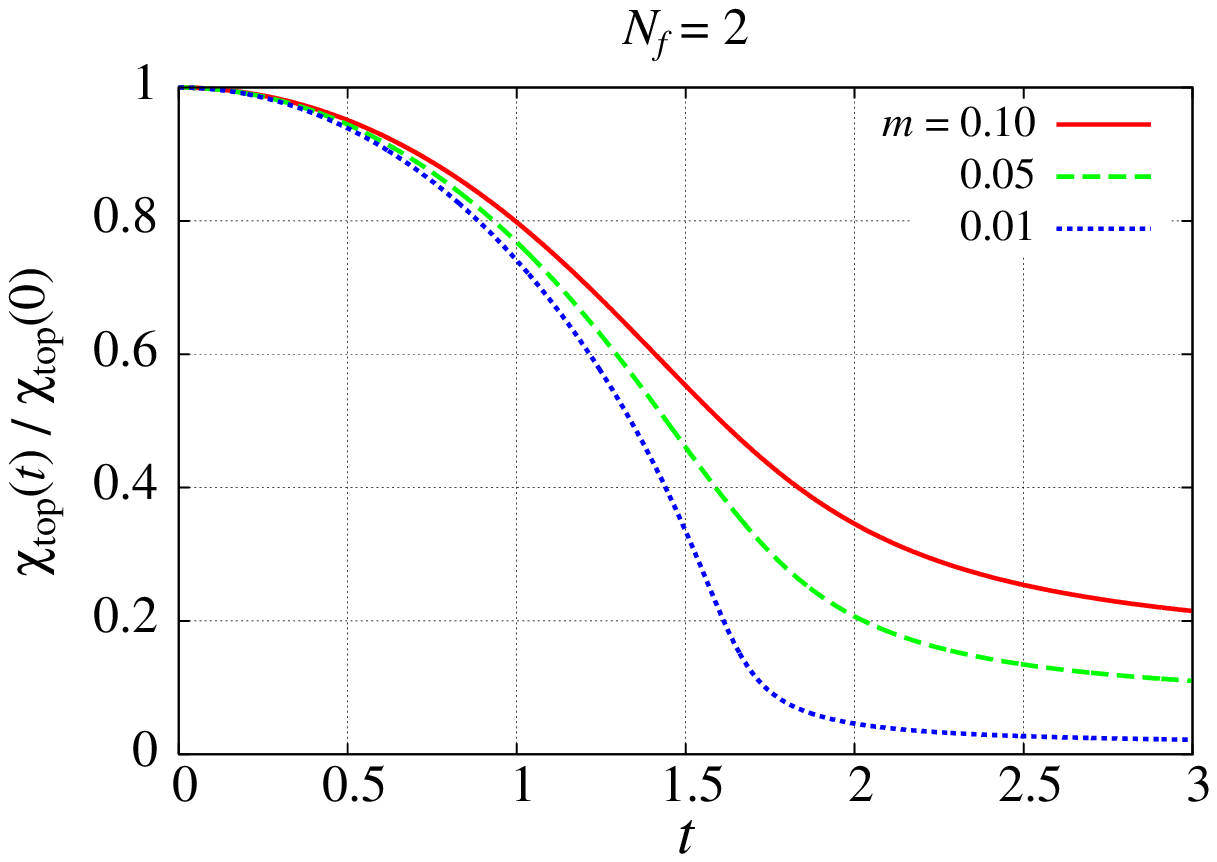}\hspace{2.5em}
\includegraphics[width=8cm]{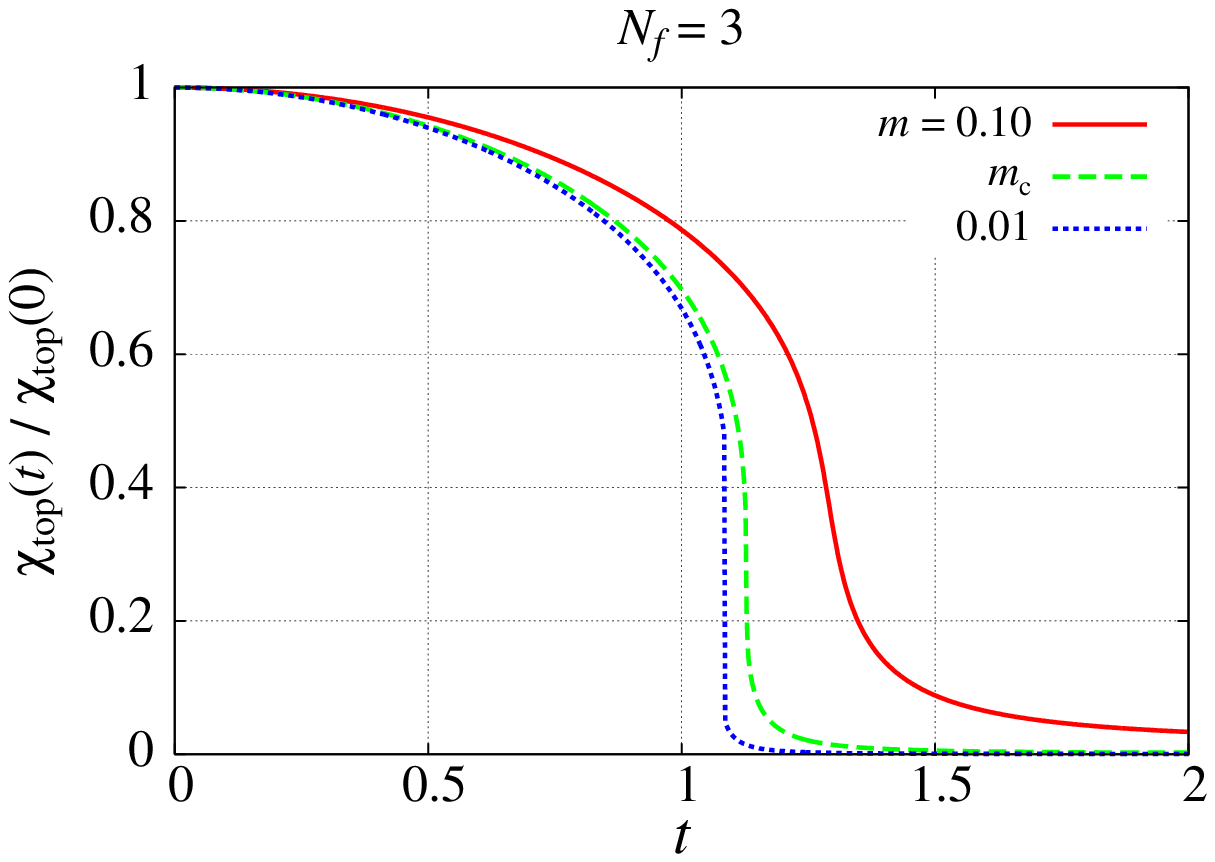}
 \caption{Temperature dependence of 
topological susceptibilities for $N_f$=2 (left) and 3 (right).
\label{fig:top}
}
\end{center}
\end{figure}

\section{Discussions and summary}
\label{sec:IV}

In this work we have considered the ChRM model with the near-zero and
topological zero modes. 
It was known before that summation over the topological zero modes 
with the Poisson distribution results in the determinant interaction, 
but unfortunately it gives a pathologic unbound potential for
the ChRM model with $N_f=3$\cite{JanikNZ97}.
On the other hand, the Gaussian distribution for topological charge $\nu$
with the total number of the modes fixed,
leads to the log-determinant type interaction~(\ref{eq:GaussSum}). 
This term in the potential resolves the \UA(1) problem at zero and finite
temperatures as suggested in Ref.~\cite{OhtaniLWH08,LehnerOVW09}, 
but still yields a second-order phase transition at finite temperature,
irrespective of $N_f$.

We have proposed that (i) the numbers of the topological zero modes with 
right and left chiralities, respectively, have the distributions
related to instanton dynamics in a finite volume, and thus
(ii) these distributions have an upper bound of ${\cal O}(N)$. 
We have adopted the binomial distribution as such a distribution.
This gives rise to the stable potential which 
describes the chiral phase transitions of the second and the first order
depending on the number of flavors
$N_f$=2 and 3, respectively, and resolves the \UA(1) problem as well.
We have confirmed through numerical evaluations that
the proposed model also reasonably reproduces temperature dependence
of meson masses, (pseudo-) scalar susceptibilities and
topological susceptibility.
Notably, the topological susceptibility 
is consistent with the universal quark mass dependence,
satisfying the anomalous Ward identity.

In the conventional ChRM models, the near-zero modes themselves
are customarily assumed to emerge from the instanton dynamics, in
contrast to 
the explicit separation of the near-zero modes and the topological zero
modes in our modeling. 
To assess the foundations and relations of these models,
a rigorous analysis of these low-lying modes based on the microscopic
dynamics would be needed, which is beyond the scope of our current study.

We have adopted the independent binomial distribution for $N_\pm$ 
as a simple improvement from the Poisson distribution in order to obtain
a stable effective potential. However, the distributions of
$N_\pm$ can be correlated non-trivially in general.  
It will be interesting to examine other possibilities for number 
distributions of the topological zero modes.
Incidentally, we note here that 
the determinant interactions of the from (\ref{eq:njlpf})
in the NJL models in 3+1 dimensions also
lead to the unbound effective potentials
for a large quark condensate beyond the cutoff although the local
minimum is usually chosen as the physical ground state.

Concerning phenomenological applications, first we need to tune
the model parameters $\Sigma$, $\alpha$ and $\gamma$ as well
as the light and strange quark  masses $m_{\rm ud}$ and $m_{\rm s}$ 
so as to reproduce the empirical properties in the vacuum. 
Extension to the case at finite baryo-chemical
potential is straightforward\cite{Stephanov:1996ki,Halasz:1998qr},
which allows us to study the phase diagram of the ChRM model with the
flavor dependence  in the space of the temperature, the 
chemical potential and the quark masses. 
Especially the existence of the analog of the QCD critical point(s)\cite{QCDCP}
will be an important subject to be studied.  
Application of this model at finite isospin and strangeness
chemical potentials are also planned.
Progress in this direction will be reported elsewhere.

\acknowledgments
The authors are grateful to T.~Wettig for fruitful discussions on this work.
M.O.\ thanks T.~Hatsuda, C.~Lehner and J.J.M.~Verbaarschot
for collaboration on a related subject.
H.F.\ and T.S.\ acknowledge useful discussions in
the workshop on
``Non-equilibrium quantum field theories and dynamic critical phenomena''
at Yukawa Institute of Theoretical Physics in March 2009.
They are also very grateful to the members of Komaba theory group for
their interests and encouragements.
This work is supported in part by Grants-in-Aid (\# 19540273) of MEXT,
Japan.

\appendix

\section{Trinomial Distribution}
Here we deal with the trinomial distribution for
$N_+$ and $N_-$, as a simplest example of the non-factorizable distribution:
\begin{align}
P(N_+, N_-)
=
\frac{(\gamma N) !}{
N_+ ! N_- ! (\gamma N-N_+ -N_-)!}
\;
p_+^{N_+} p_-^{N_-}
(1-p_+-p_-)^{\gamma N -N_+ -N_-}
\; , 
\label{eq:tri_dist}
\end{align}
where $p_{+(-)}$ is the probability for a single cell to be occupied
by an (anti-)instanton.
Note that the distribution is symmetric under the exchange of $+$ with $-$. 
Replacing $P(N_+)P(N_-)$ in eq.~(\ref{eq:ia_sum}) with $P(N_+,N_-)$
and setting $p_+ = p_- = p$, we obtain the effective potential
\begin{align}
\Omega_{\rm Tri}(S;m,\theta,t)
=
\half\Sigma^2 {\rm tr} S^\dagger S
-
\half\ln \det 
        \left[(S+{\cal M})(S^\dagger +{\cal M}^\dagger )+t^2 \right]
-\half 
        \gamma \ln \left[\alpha e^{i\theta}\det(S+{\cal M})
        +\alpha e^{-i\theta}\det(S^\dagger +{\cal M}^\dagger)+1\right]
\; 
\label{eq:tri_pot}
\end{align}
with $\alpha = p/(1-2p)$. 
This is quite similar to eq.~(\ref{eq:thepf}), and qualitative
features of the model are unchanged.
The quenched topological susceptibility, or the variance of $\nu=N_+ -N_-$, for eq. (\ref{eq:tri_dist}) 
is computed as
\begin{align}
\tau
=
\gamma p
\; .
\end{align}

\end{document}